\documentclass[%
prb,%
twocolumn,
aps,%
showpacs,%
a4paper,%
superscriptaddress,%
floatfix%
]{revtex4-1}
\usepackage{CJKutf8}
\usepackage{graphicx}
\usepackage{bm}
\usepackage{amsmath}
\usepackage{amssymb}
\usepackage{float}
\usepackage[utf8]{inputenc}
\usepackage[T1]{fontenc}
\usepackage[USenglish]{babel}
\usepackage{microtype}
\usepackage{nicefrac}
\usepackage{booktabs}
\usepackage{xspace}
\usepackage{subcaption}
\usepackage[color=black!10]{todonotes}

\usepackage[colorlinks,citecolor=blue,urlcolor=blue,linkcolor=blue,bookmarks=false,filecolor=blue,runcolor=blue]{hyperref}
\usepackage[all]{hypcap}
\newcommand{\figref}[1]{Fig.~\ref{#1}}

\renewcommand{\Im}{%
\mathrm{Im}
}

\newcommand{\rangstrom}[1]{
\ensuremath{#1\,\text{\AA}^{-1}}
}
\newcommand{\wse}{WSe$_2$\xspace}
\newcommand{\mos}{MoS$_2$\xspace}
\newcommand{\tmdcs}{TMDCs\xspace}
\begin{document}
\begin{CJK*}{GB}{}
\title{Non-Generic Dispersion of Excitons in the Bulk of  \wse}
\author{R. Schuster}
\author{Y. Wan}%\author{Y. Wan (万宇)}
\author{M. Knupfer}
\affiliation{IFW Dresden, Institute for Solid State Research, P.O. Box 270116, D-01171 Dresden, Germany}
\author{B. Büchner}
\affiliation{IFW Dresden, Institute for Solid State Research, P.O. Box 270116, D-01171 Dresden, Germany}
\affiliation{Department of Physics, Technische Universität Dresden, 01062 Dresden, Germany}

\date{\today}

\begin{abstract} 
  We combine electron energy-loss spectroscopy (EELS) and density
  functional theory (DFT) calculations to study the dispersion and
  effective mass of excitons in the bulk of \wse. Our EELS data
  suggest substantial deviations from the generic quadratic momentum dependence
  along the $\Gamma K$-direction. From the DFT-derived Kohn-Sham
  states we deduce the EELS response without the inclusion of
  particle-hole attraction to study the possible role of the
  single-particle band structure on the exciton behavior. Based on
  this analysis we argue in favor of a strongly momentum dependent
  particle-hole interaction in \wse and other group VI-transition-metal
  dichalcogenides.
\end{abstract}
%\pacs{71.20.Mq,71.45.Gm,78.20.Ci,79.20.Uv}
\maketitle
\end{CJK*}
%\listoftodos
%\rstodo{check pacs numbers}
\section{Introduction}
Transition-metal dichalcogenides (\tmdcs) in general and the
semiconducting group-VI representatives in particular have been
investigated for decades \cite{Wilson_Adv.Phys._1969}. While
originally of interest for their prototype role as anisotropic,
quasi-2D semiconductors with indirect band gaps and their chemical and
structural similarities to the metallic group V-relatives with
charge-ordered and/or superconducting ground states, \mos and related
compounds like \wse (re)gained substantial attention after it was
realized that they exhibit a crossover from an indirect to a direct
band gap as a function of layer number
\cite{Mak_Phys.Rev.Lett._2010_v105_p136805} yielding exceptionally
large quantum yields in their photoluminescence  spectra. This
discovery can be considered a milestone for the field of
optoelectronics in 2D semiconductors and it sparked a lot of research
related to possible applications. In addition, it is by now well
established that the fundamental physics of excitons – which are
at the heart of the optical properties of \tmdcs – is
extraordinarily rich. This is related (i) to the honeycomb structure
of the individual layers and (ii) the strong spin-orbit coupling (SOC)
of the constituent atoms. By symmetry this dictates several remarkable
observations, among them a finite contribution of the Berry curvature to
the exciton dynamics
\cite{Zhou_Phys.Rev.Lett._2015_v115_p166803,Srivastava_Phys.Rev.Lett._2015_v115_p166802}
and the so called valley degree of freedom
\cite{Yu_Natl.Sci.Rev._2015_v2_p57,Xu_NatPhys_2014_v10_p343}.

While much is known about absorption and emission processes of \tmdcs in the optical limit, i.e., the case of vertical transitions between occupied and unoccupied states \cite{Zeng_Chem.Soc.Rev._2015_v44_p2629,Matsud_J.Phys.Soc.Jpn._2015_v84_p121009}, the response to probes allowing for non-vertical transitions between the valence- and conduction-states represents a widely open field. Here we employ a combination of electron-energy loss spectroscopy (EELS) in transmission together with first-principle calculations based on density-functional theory (DFT) to gain insight into the momentum dependence of exciton absorption in the bulk of \wse. This provides access to e.g. the effective mass of excitons and the $q$-dependence of particle-hole interactions.

\section{Experiments and Results}
Electron energy-loss spectroscopy (EELS) in transmission
\cite{Fink_Adv.Electron.ElectronPhys._1989} is a bulk-sensitive
scattering technique whose cross-section is proportional to the so
called loss-function
$L(\bm{q},\omega)=\Im(-1/\epsilon(\bm{q},\omega))$, with
$\epsilon(\bm{q},\omega)$ the momentum- and energy-resolved dielectric
function \cite{Sturm_Z.Naturforsch._1993}. It has been successfully
applied to investigate collective charge modes in numerous
condensed-matter systems (see
\cite{Roth_J.ElectronSpectrosc.Relat.Phenom._2014_v195_p85,Fink_J.ElectronSpectrosc.Relat.Phenom._2001_v117-118_p287}
for an overview). Recently, the momentum dependence of excitons in
other topical quasi-2D materials like \mos
\cite{Habenicht_Phys.Rev.B_2015_v91_p245203} and black phosphorus
\cite{Schuster_Phys.Rev.Lett._2015_v115_p26404} has been successfully
investigated with the same methodology. For the present experiments
bulk-like thin ($d\approx 100~\text{nm}$) films of \wse have been
prepared by exfoliation from a single crystal purchased from ``2D
semiconductors'' and measured in a purpose-built EELS spectrometer
\cite{Fink_Adv.Electron.ElectronPhys._1989} with energy and momentum
resolutions of $\Delta E\approx90~\text{meV}$ and $\Delta q\approx\rangstrom{0.03}$,
respectively. All data were taken at a temperature of about
$20~\text{K}$ to reduce thermal broadening.

%\subsection{Exciton Dispersion}
%\label{sec:exciton-dispersion}

\begin{figure}[ht]
  \centering
  \includegraphics[scale=.75]{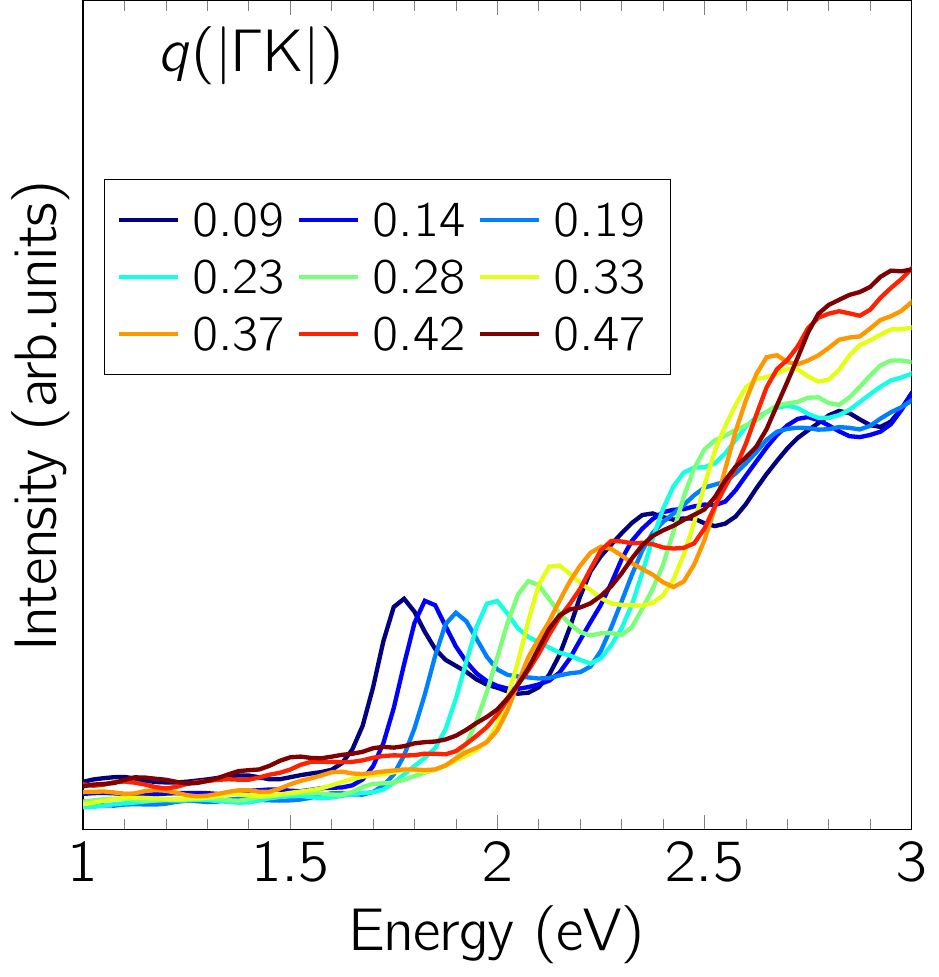}
  \caption{(Color online) Momentum dependent EELS intensity of bulk \wse along $\Gamma K$.}
  \label{fig:momentum_dep}
\end{figure}

In \figref{fig:momentum_dep} we show the momentum dependent EELS intensity along the $\Gamma K$ line of the hexagonal Brillouin zone (BZ).

It is well known that \tmdcs exhibit a prominent splitting of the
valence band (VB)-edge around the $K$-point, the origin of which is to
some extend a matter of controversy but is believed to originate from  SOC for
the monolayer (ML) and from a combination of  SOC and inter-layer
hybridization for few-layer and bulk systems
\cite{Molina-Sanchez_Phys.Rev.B_2013_v88_p45412}. The size of this
splitting varies for different representatives and has been calculated
to about $0.5\,\text{eV}$ for \wse
\cite{Roldan_Ann.Phys._2014_v526_p347}. This value is confirmed
experimentally by angle-resolved photoemission
\cite{Riley_NatPhys_2014_v10_p835} and our own calculations (see below). The spectral shape of the optical
response in \tmdcs is a consequence of the superimposed
joint-densities of states for the two VBs at the $K$-point with the
conduction band (CB) minimum
\cite{Molina-Sanchez_Phys.Rev.B_2013_v88_p45412}. On top of these two
step-function-like continua, prominent excitonic features emerge due
to sizable particle-hole attraction which are known to be strongly
enhanced upon approaching the ML limit. Compared to \mos
\cite{Habenicht_Phys.Rev.B_2015_v91_p245203} the higher value of the
SOC at the $K$-point and a suppressed single-particle continuum,
enhances the visibility of the exciton features of \wse and allows to
trace the momentum dependent exciton absorption up to much higher
$q$-values. Consequently, the data shown in \figref{fig:momentum_dep}
suggest a substantial exciton dispersion to be discussed below.

\begin{figure}[ht]
  \centering
  \includegraphics[scale=.75]{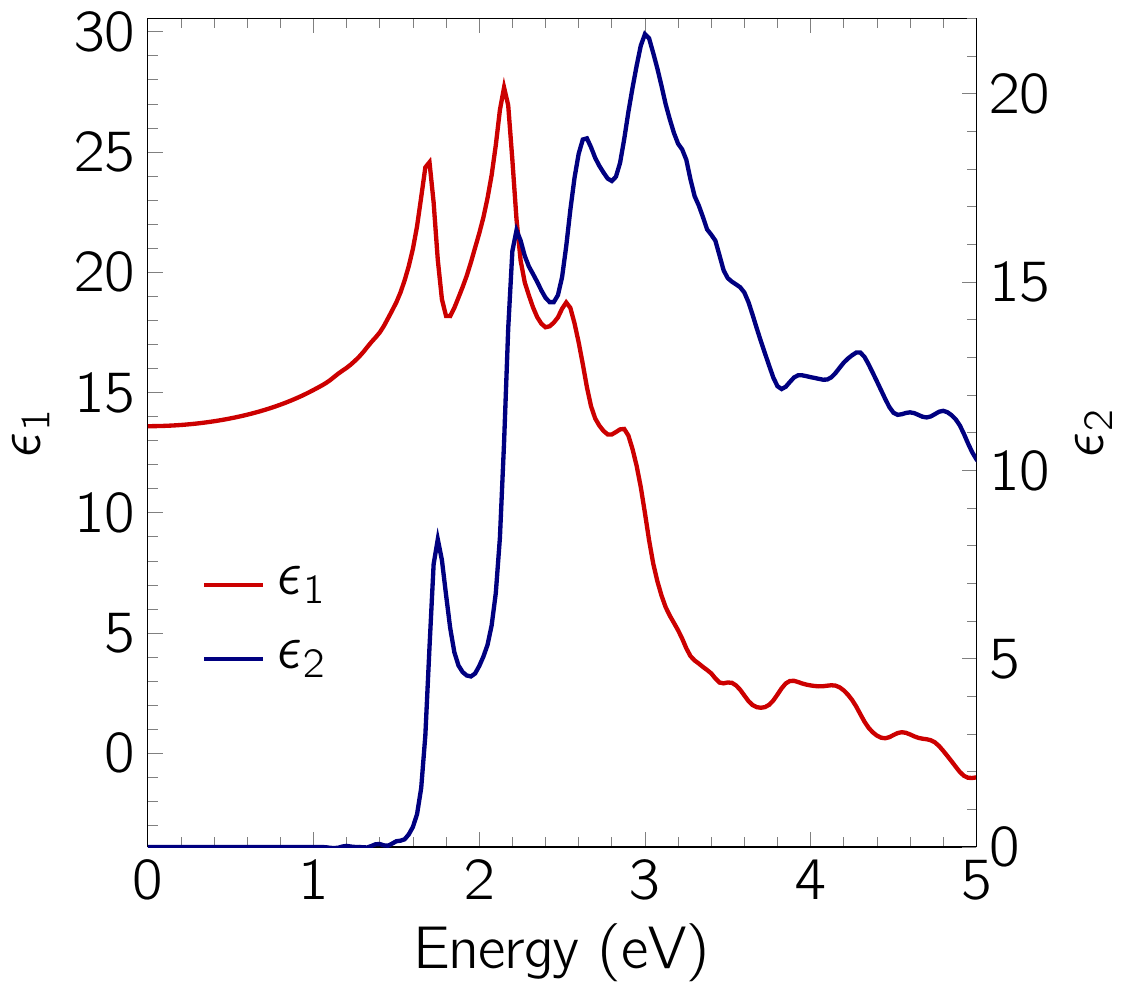}
  \caption{(Color online) Dielectric function for \wse as obtained
    from a KKA of the measured EELS intensity. The first absorption
    peak lies at 1.75~eV which is in excellent agreement with previous
    reports
    \cite{Beal_J.Phys.C._1976_v9_p2459,Anedda_Can.J.Phys._1979_v57_p368}.}
  \label{fig:optical_functions}
\end{figure}

To retrieve the complex dielectric function
$\epsilon(\omega)=\epsilon_1(\omega)+\mathrm{i}\epsilon_2(\omega)$,
EELS data have been corrected for contributions from the elastic line
and multiple scattering and normalized to yield $\epsilon_1(\omega=0)$
in agreement with earlier reports
\cite{Beal_J.Phys.C._1976_v9_p2449}. The resulting curves for the real
and imaginary part of the dielectric function are shown in
\figref{fig:optical_functions}.

Note that the first peak in the absorption spectrum (proportional to
$\epsilon_2(\omega)$) occurs at $1.75~\text{eV}$. Given that the
calculated single-particle gap at the GW-level amounts to about
$1.75~\text{eV}$ \cite{Jiang_J.Phys.Chem.C_2012_v116_p7664} the
exciton binding energy of bulk-\wse is very small
\cite{Anedda_Can.J.Phys._1979_v57_p368} which makes an interpretation
of the observed particle-hole pair in terms of a Wannier-Mott like
state appealing.  This behavior is to be contrasted with the
prediction of a much higher exciton binding energy of $0.9~\text{eV}$
in the ML of \wse
\cite{Ramasubramaniam_Phys.Rev.B_2012_v86_p115409}. This is supported
by calculations for the spread of the excitonic wavefunction in the
bulk of the related \mos-system
\cite{Molina-Sanchez_Phys.Rev.B_2013_v88_p45412}. Note, however that
the coincidence between excitonic binding energy and the corresponding
level of (de)localization is ambiguous for the case of MLs as in the
2D limit screening is strongly reduced which always yields a sizable
exciton binding energy, independent of the spatial extension of the
particle-hole pair \cite{Cudazzo_Phys.Rev.Lett._2016_v116_p66803}.

\begin{figure}[ht]
  \centering
  \includegraphics[scale=.75]{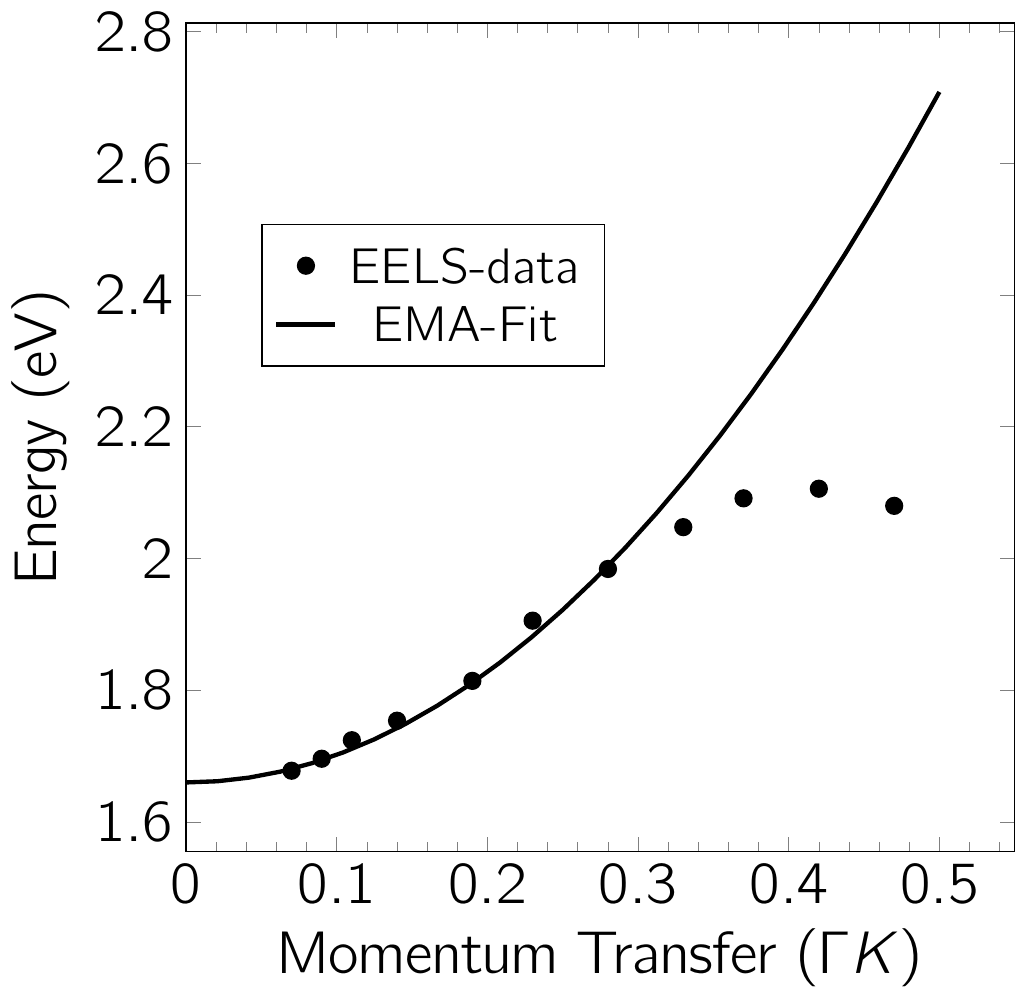}
  \caption{The dispersion of the lowest energy structure in the data shown in \figref{fig:momentum_dep} together with corresponding low-$q$ fits of the generic form $E(q)=E_0+A\cdot q^2$. From the fit we obtain an effective exciton mass of $M\approx0.91~m_0$. Note the marked discrepancy for higher momentum values and in particular the downturn of the data points for $q\sim 0.4\left|\Gamma K\right|$.}
  \label{fig:dispersion}
\end{figure}

In the case of a weakly-bound Wannier-Mott exciton, the dispersion of
the two-particle wavepacket is given within the so called effective
mass approximation (EMA) \cite{Yu_FundamentalsSemiconductors_2010} by

\begin{equation}
  \label{eq:1}
  E(q)=\frac{\hbar^2q^2}{2M^\ast}+E_0\,
\end{equation}
with the effective exciton mass given as the sum of the effective
electon- and hole-mass-values $M^\ast=m_e^\ast+m_h^\ast$. Although it
has been shown recently that the EMA must be augmented to capture the
contribution from a finite Berry-curvature
\cite{Zhou_Phys.Rev.Lett._2015_v115_p166803,Srivastava_Phys.Rev.Lett._2015_v115_p166802}
for excitonic states with non-zero angular momentum, we emphasize that
our energy resolution is not sufficient to observe excited states in
the Rydberg series and we take Eq.~\eqref{eq:1} to describe our
experimental data shown in
\figref{fig:momentum_dep}. \figref{fig:dispersion} contains the
experimentally observed exciton dispersion as obtained from the
low-energy edges of the data shown in \figref{fig:momentum_dep}
together with a low $q$-fit according to the EMA. As can be seen, in
the vicinity of the BZ center the EMA yields excellent agreement with
the experimental data with a derived exciton mass of
$M^\ast\approx0.91~m_0$ with the free electron mass $m_0$. In contrast to
this, for higher $q$-values there is an increasing discrepancy between
the model and the data. In particular, we find a downturn of the
dispersion at $q\sim 0.4\left|\Gamma K\right|$ which
is in stark contrast to the monotonic behavior predicted within the
EMA.

Deviations from the generic quadratic momentum dependence of the
exciton dispersion are well expected in the case of degenerate
bands. It was realized early on that the canonical coordinate
transformation to solve the Schrödinger equation for the two-particle
state is not possible in the case of degenerate bands
\cite{Dresselhaus_J.Phys.Chem.Solids_1956_v1_p14}. This circumstance
is known to yield deviations from the simple parabolic exciton band
structure \cite{Altarelli_Phys.Rev.B_1977_v15_p4898}. While the
valence band edges at the $K$-point of the hexagonal BZ are strongly
split because of the strong SOC, the CB minima are nearly (cf. our
band structure calculations discussed below) degenerate at the same
point in the BZ. But even if this degeneracy was present, it is lifted
already for small deviations from the high-symmetry point, and we
therefore exclude this effect as being responsible for the
non-parabolic shape observed in
\figref{fig:dispersion}. 

\begin{figure}[ht]
  \centering
  %\begin{subfigure}{.425\textwidth}
    \includegraphics[scale=.3]{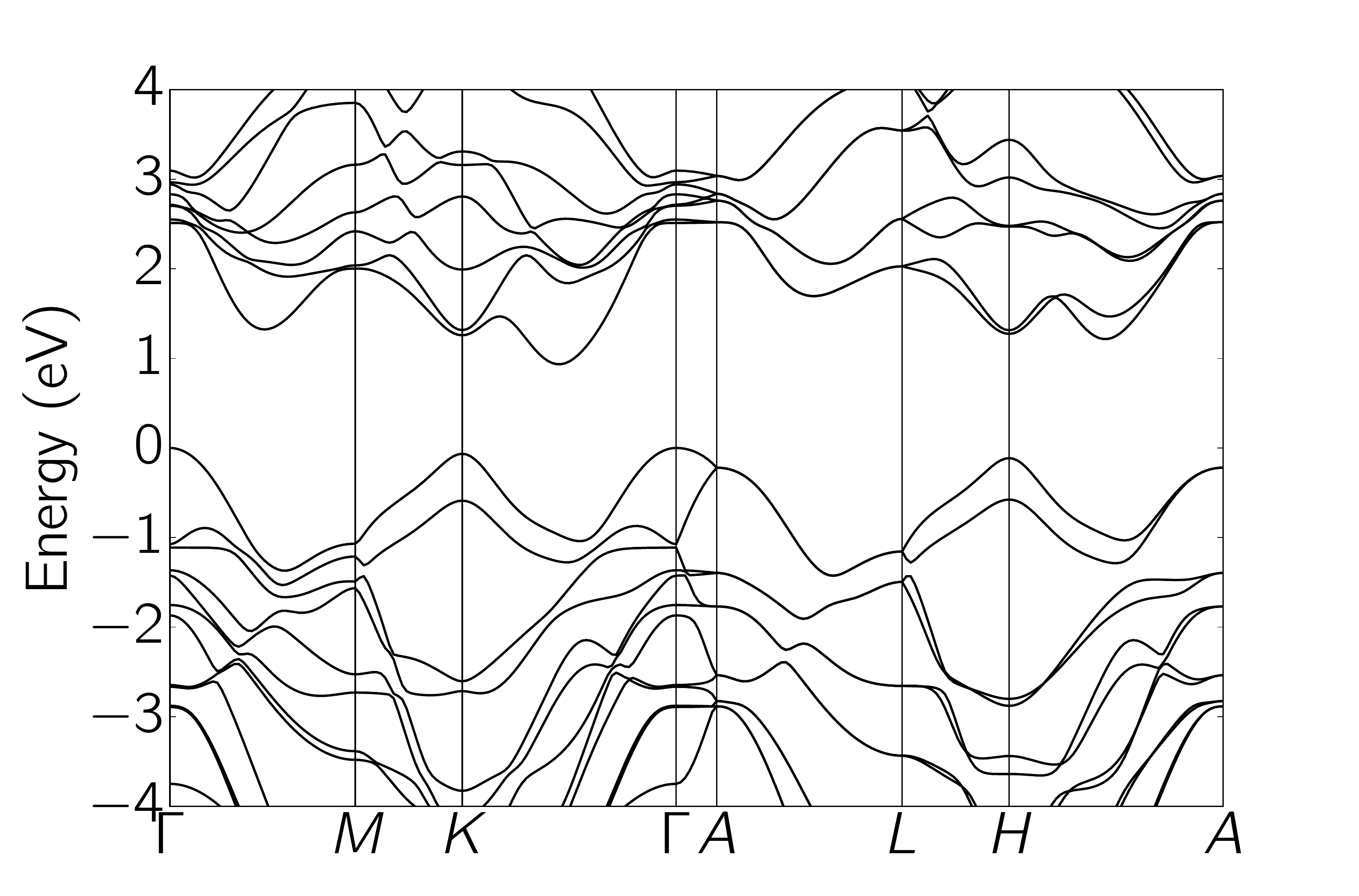}
  %\end{subfigure}%\hfill
  %\begin{subfigure}{.425\textwidth}
  %  \includegraphics[scale=.25]{fig4b}
  %\end{subfigure}
  \caption{GGA-derived single-particle band structure for \wse. The zero of the energy axis was set to the VB maximum at the $\Gamma$-point.}
  \label{fig:band_structure}
\end{figure}

Another possible source of a non-generic exciton momentum dependence
might be an interference with the absorption caused by interband
transitions. To investigate this issue we have performed
first-principle calculations with the FPLO package
\cite{Koepernik_Phys.Rev.B_1999_v59_p1743} including SOC and employing
the generalized gradient approximation (GGA) as parametrized by
Perdew, Burke and Ernzerhof
\cite{Perdew_Phys.Rev.Lett._1996_v77_p3865}.  As input we used the
crystal structure given in
\cite{Schutte_J.SolidStateChem._1987_v70_p207} except for the in-plane
lattice constant which we derived \emph{in-situ} from our own electron
diffraction data. The total density was converged on a grid of
$24\times24\times12$ irreducible $k$-points. The calculated
single-paricle band structure is shown in \figref{fig:band_structure}.

In agreement with previous reports \cite{Roldan_Ann.Phys._2014_v526_p347,Jiang_J.Phys.Chem.C_2012_v116_p7664}, our DFT calculations yield an indirect gap of
$0.93\,\text{eV}$, the VB maximum at the $\Gamma$-point (lying
$\sim70\,\text{meV}$ above the maximum at the $K$-point), a
substantial ($\sim520\,\text{meV}$) SO-induced splitting of the VB
around the $K$-point and the CB-minimum approximately halfway along the
$\Gamma K$-line. The  values for the effective masses derived
from our calculations are given in Tab.~\ref{tab:masses}.

\begin{table}[ht]
  \centering
\caption{Effective mass values derived from a quadratic expansion of the Kohn-Sham states shown in  \figref{fig:band_structure} around the $K$-point. All values are in units of the free electron mass $m_0$.}
  \begin{tabular}{lc}
    \toprule
    Quasiparticle&Effective Mass\\
    \midrule
    electron ($m^\ast_e$)&0.548\\
    light hole ($m^\ast_{LH}$)&0.465\\
    heavy hole ($m^\ast_{HH}$)&0.607\\
    exciton ($M^\ast=m^\ast_e+m^\ast_{LH}$)&1.01\\
    %$M_B=m_e+m_{HH}$&1.16\\
    \bottomrule
  \end{tabular}
  \label{tab:masses}
\end{table}

As can be seen, our DFT calculations \emph{over}estimate the exciton
effective mass derived from our EELS data shown in
\figref{fig:dispersion}. The origin of this discrepancy is unknown at
the moment. Usually, DFT tends to underestimate effective mass values
which can be improved by quasiparticle corrections in terms of more
elaborate GW-calculations. However, from corresponding data for the
\tmdcs it is known that the curvature of the Kohn-Sham derived band
structure (in particular those from the VB region) are stable against
the self-energy corrections induced by the GW-term
\cite{Jiang_J.Phys.Chem.C_2012_v116_p7664,Molina-Sanchez_Phys.Rev.B_2013_v88_p45412}. Therefore
effective masses obtained within DFT are supposedly close to the
ones obtained from GW-renormalized bands.

\begin{figure}[ht]
  \centering
%  \begin{subfigure}{.425\textwidth}
    \includegraphics[scale=.35]{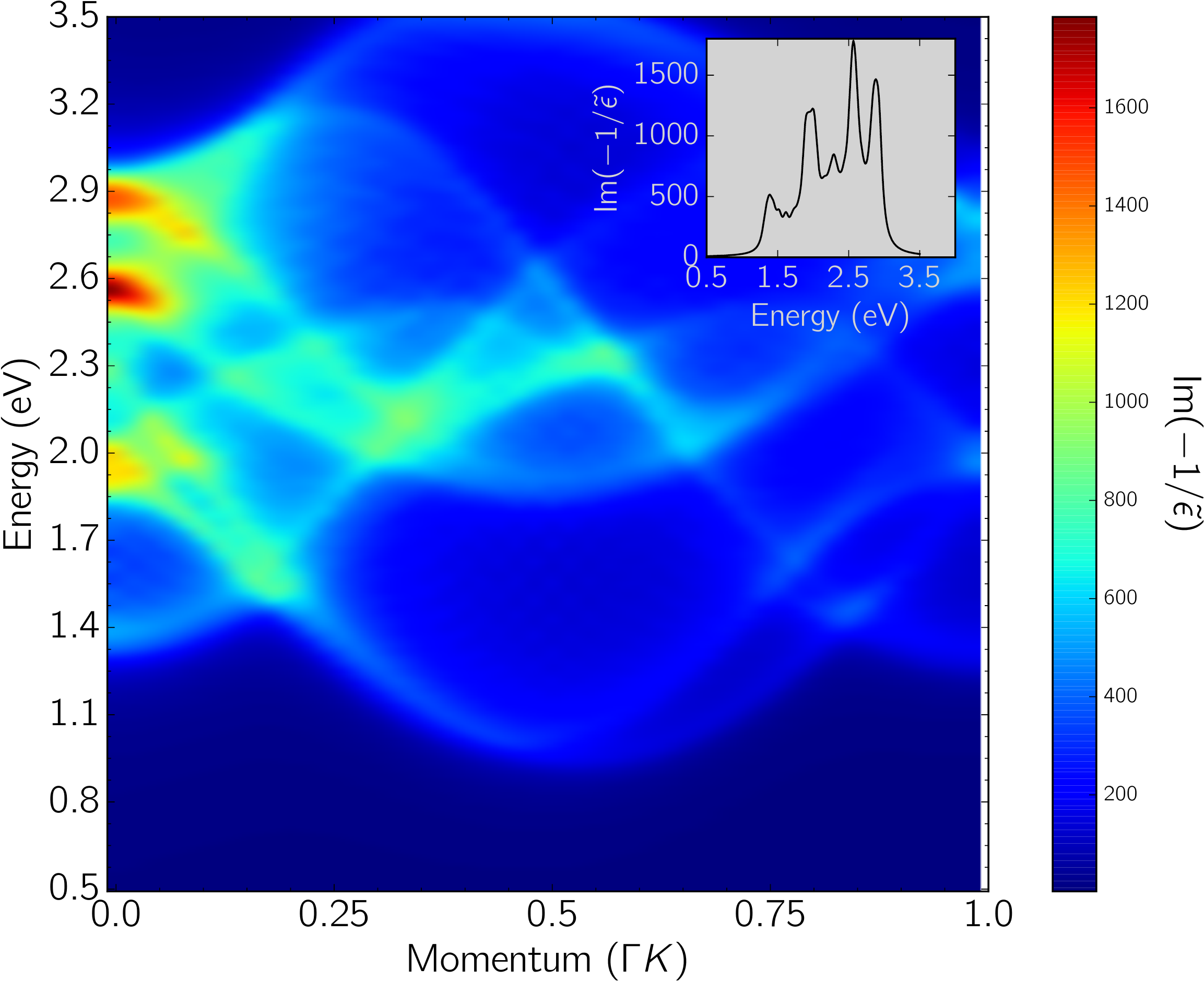}
    %\phantomcaption
%    \label{fig:fig5a}
%  \end{subfigure}
\\%\hspace*{2em}
%  \begin{subfigure}{.425\textwidth}
    \includegraphics[scale=.275]{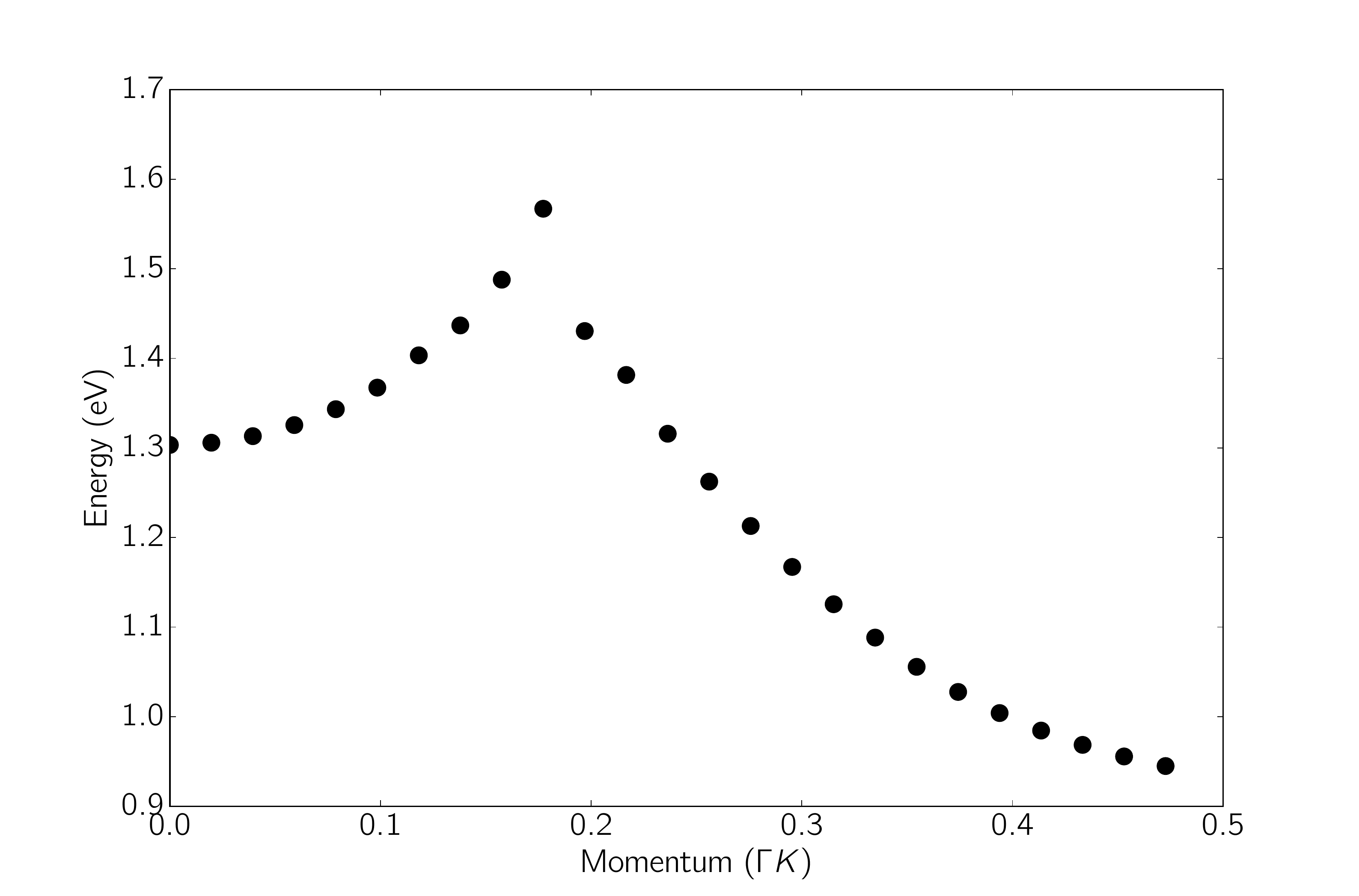}
    %\phantomcaption
%    \label{fig:fig5b}
%  \end{subfigure}
  \caption{(Color online) Top Panel: $\text{Im}(-1/\tilde{\epsilon}(q,\omega))$ as calculated from the simplified model along the $\Gamma K$ line according to Eq.~\eqref{eq:3}. A phenomenological broadening of $\delta=50\,\text{meV}$ has been employed. The inset shows $\text{Im}(-1/\tilde{\epsilon}(0,\omega))$. Bottom panel: The dispersion obtained from the edge of the lowest-energy feature. Note the prominent maximum at $q\sim 0.18~\left|\Gamma K\right|$}
  \label{fig:dft_lf}
\end{figure}

To investigate the possible effect of interband transitions on the exciton dispersion we define a simplified one-dimensional (along the $\Gamma K$-cut) model for the susceptibility
\begin{equation}
  \label{eq:2}
  \chi_0(q,\omega)=2\,\sum_{v,c,k}\;\frac{1}{\omega-(\epsilon^c_{k+q}-\epsilon^v_k)+\text{i}\delta}
\end{equation}
from which we can derive an effective dielectric function according to
\begin{equation}
  \label{eq:3}
  \frac{1}{\tilde{\epsilon}(q,\omega)}=1+\frac{1}{q^2}\,\chi_0(q,\omega)\,.
\end{equation}
Here, the $\epsilon_k^{c(v)}$ correspond to conduction
(valence) Kohn-Sham bands derived from our GGA-calculations shown in
\figref{fig:band_structure}, $q,k\in(0,\left|\Gamma K\right|)$ and the
sum runs over all VB and CB states (except for
spin-degeneracy which causes the factor of two) up to a defined energy
cut-off ($3.5\,\text{eV}$ in the present case). The reason for the
latter is to keep transitions originating from particular $q$-points
discernible. If the cut-off is chosen too high, all features become
smeared in a broad continuum preventing a straightforward mode
assignment.  This model contains two essential assumptions: (i) The
excitonic bands follow the Kohn-Sham single-particle bands
$\epsilon_k^{c(v)}$, i.e. we neglect particle-hole
attraction. While this appears unphysical at first sight, we justify
this by the very small exciton binding energy as discussed above. (ii) The matrix element mediating the transitions between the VB and
the CB, which in a more realistic model would enter the numerator of
Eq.~\eqref{eq:2}, is constant for $q\|\Gamma
K$. %In principle it would be possible to calculate this matrix element by projecting the Kohn-Sham states onto Wannier-functions but this is beyond the scope of the present work.
This is motivated by the observation that the orbital weight of the VB
and CB states in TMDCs vary only moderately along the $\Gamma
K$-direction \cite{Cappelluti_Phys.Rev.B_2013_v88_p75409}. The result
of our model calculation is shown in \figref{fig:dft_lf}.

From the inset of \figref{fig:dft_lf} for $q\rightarrow0$ one can
identify several prominent structures in the two-particle
spectrum. Therein the lowest (highest) lying structures at energies
$E_{\text{I}}\approx1.34\,\text{eV},\;(E_{\text{II}}\approx2.53\,\text{eV})$
correspond to vertical transitions in the $K(\Gamma)$-point region of the BZ.  In the
following we are only concerned with the former. Compared to the
experimental data shown in \figref{fig:optical_functions} its
calculated energy position lies at much lower energies. This is a
consequence of the well known underestimation of the band gap in DFT
like calculations. As can be seen from \figref{fig:dft_lf}, upon
increasing the momentum transfer, this feature first disperses
quadratically to higher energies until $q\sim0.18\,\Gamma K$ where it
starts to hybridize with a downward dispersing mode which leads to an
effective red-shift of the lowest feature. The origin of this downturn
lies in the local maximum of the lowest CB along the $\Gamma K$-line
(cf.  \figref{fig:band_structure}). While it is tempting to identify
this redshift with the corresponding experimental observation shown in
\figref{fig:dispersion} we emphasize that there is a large mismatch in
the momentum location of these
maxima.  

We therefore also exclude interband transitions as the cause for the
peculiar momentum dependence shown in
\figref{fig:dispersion}. Consequently our data are suggestive of a
strong momentum dependence of the particle-hole attraction which is
supposed to renormalize the denominator of the two-particle response
function in Eq.~\eqref{eq:2}. Indeed, recently the momentum dependence
of the exchange interaction has been shown to play a prominent role in
the interpretation of the exciton character in MLs of \tmdcs
\cite{Cudazzo_Phys.Rev.Lett._2016_v116_p66803}. In particular, it has
been argued that the momentum dependence of the exchange interaction
leads to a flattening of the exciton band structure away from the zone
center if the exciton wave function is localized in the limit
$q\rightarrow 0$. From our observation of a strongly non-generic
exciton dispersion in the bulk of \wse we are led to speculate that
already in the bulk-limit the interaction of the particle-hole pair is
non-trivial. \wse provides an ideal platform to study this behavior as
the enhanced SOC and the overall shape of the interband continuum
allows to track the exciton up to much higher momentum values compared
to the related \mos.

\section{Summary}
\label{sec:summary}
In summary, we investigated the dispersion of excitons in the bulk of
\wse with EELS. In the vicinity of the BZ center we observe a
quadratic momentum dependence which is the generic behavior expected
for a Wannier-Mott like state. Upon increasing the momentum transfer
there are, however, substantial deviations from this picture.  From a
comparison with the optical response generated by uncorrelated
interband transitions calculated from density-functional theory we
argue in favor of a non-trivial particle-hole interaction in the bulk
of \tmdcs.

\section{Acknowledgments}
We highly appreciate technical support from R. Hübel, S. Leger and M. Naumann and helpful guidance from T. Ritschel concerning the DFT calculations.
\bibliographystyle{apsrev4-1}
\bibliography{exciton_dispersion_wse2_manuscript}
%\bibliography{/home/roman/work/paper_drafts/exciton_dispersion_wse2/manuscript,/home/roman/literature/cdw,/home/roman/literature/general_theory,/home/roman/literature/eels,/home/roman/literature/textbooks,/home/roman/literature/mypapers,/home/roman/literature/phosphorene}
%\listoftodos
\end{document}